\documentclass[aps,prl,twocolumn,groupedaddress,amssymb,superscriptaddress]{revtex4-1}
\usepackage[usenames]{color}
\usepackage[normalem]{ulem}    
\definecolor{Blue}{rgb}{0.1,0,0.9}
\definecolor{Red}{rgb}{0.9,0,0.1}
\definecolor{dg}{rgb}{0,0.5,0} 
\usepackage[dvips]{graphicx}
\DeclareGraphicsExtensions{.eps,.ps}
\usepackage{bm}      

\newcommand{\sfaof}{Sm\-Fe\-As\-O$_{0.85}$\-F$_{0.15}$}
\newcommand{\smfraof}{Sm\-Fe$_{1-x}$\-Ru$_{x}$\-As\-O$_{0.85}$\-F$_{0.15}$}
\newcommand{\refraof}{RE\-Fe$_{1-x}$\-Ru$_{x}$\-As\-O$_{0.85}$\-F$_{0.11}$}
\newcommand{\refrao}{RE\-Fe$_{1-x}$\-Ru$_{x}$\-As\-O}
\newcommand{\old}[1]{}

\begin{document}



\title{Correlated trends of coexisting magnetism and superconductivity \\  in optimally electron-doped oxy-pnictides}




\author{S.~Sanna}
\email[]{Samuele.Sanna@unipv.it}
\affiliation{Dipartimento di Fisica ``A.\ Volta'' and Unit\`a CNISM di Pavia, I-27100 Pavia, Italy}
\author{P.~Carretta}
\affiliation{Dipartimento di Fisica ``A.\ Volta'' and Unit\`a CNISM di Pavia, I-27100 Pavia, Italy}
\author{P.~Bonf\`a}
\affiliation{Dipartimento di Fisica ``A.\ Volta'' and Unit\`a CNISM di Pavia, I-27100 Pavia, Italy}
\author{G.~Prando}
\affiliation{Dipartimento di Fisica ``A.\ Volta'' and Unit\`a CNISM di Pavia, I-27100 Pavia, Italy}
\affiliation{Dipartimento di Fisica ``E.\ Amaldi'', Universit\`a di Roma3-CNISM, I-00146 Roma, Italy}
\author{G.~Allodi}
\affiliation{Dipartimento di Fisica and Unit\`a CNISM di Parma, I-43124 Parma, Italy}
\author{R.~De Renzi}
\affiliation{Dipartimento di Fisica and Unit\`a CNISM di Parma, I-43124 Parma, Italy}
\author{T.~Shiroka}
\affiliation{Laboratorium f\"ur Festk\"orperphysik, ETH-H\"onggerberg, CH-8093 Z\"urich, Switzerland}
\affiliation{Paul Scherrer Institut, CH-5232 Villigen PSI, Switzerland}
\author{G.~Lamura}
\affiliation{CNR-SPIN and Universit\`a di Genova, via Dodecaneso 33, I-16146 Genova, Italy}
\author{A.~Martinelli}
\affiliation{CNR-SPIN Corso Perrone 24, I-16146 Genova, Italy}
\author{M.~Putti}
\affiliation{CNR-SPIN and Universit\`a di Genova, via Dodecaneso 33, I-16146 Genova, Italy}




\date{\today}

\begin{abstract}
We report on the recovery of the short-range static magnetic order and on the concomitant degradation of the superconducting state in optimally F-doped \smfraof\ for $0.1\leq x\lesssim 0.5$. The two reduced order parameters coexist within nanometer-size domains in the FeAs layers and finally disappear around a common critical threshold $x_c\sim 0.6$.
Superconductivity and magnetism are shown to be closely related to two {\em distinct} well-defined local electronic environments of the FeAs layers. The two transition temperatures, controlled by the isoelectronic and diamagnetic Ru substitution, scale with the
volume fraction of the corresponding environments. This fact indicates that superconductivity is assisted by magnetic fluctuations, which are frozen whenever a short-range static order appears, and totally vanish above the magnetic dilution threshold $x_c$.
\end{abstract}


\maketitle


The appearance of high-$T_c$ superconductivity (SC) close to the disruption of static magnetic (M) order is a
general feature of the Fe-based superconductors either as a function of doping or external pressure. In the
REFeAsO family (RE1111) it is found that SC and M strongly compete and hardly coexist simultaneously
\cite{Luetkens2009, Khasanov2011}, apart for RE=Sm and Ce \cite{Sanna2009, Sanna2010} within a small doping range where both order parameters are depressed.
Coexistence implies short range magnetic order, that is
detected only by local probes such as muon-spin rotation ($\mu$SR) \cite{Sanna2010} or nuclear quadrupole resonance (NQR)
\cite{Lang2010}, since it eludes long coherence probes such as powder diffraction \cite{Zhao2008}. The competition between the superconducting and magnetic ground-states must be reconciled with the prevailing models of pairing mediated by spin fluctuations \cite{Boeri2008}. These models are seemingly in contradiction with the evidence that the two mutually excluding orders coexist {\em only} when phase separation occurs.

Here we show, by means of $\mu$SR and $^{75}$As NQR, that magnetism is surprisingly still at play in optimally F-doped \smfraof.
The isoelectronic Fe:Ru substitution is found to deteriorate the
superconducting state in optimally electron-doped \sfaof\ samples and simultaneously to recover static magnetism within the FeAs layers, for $0.1\leq x\lesssim 0.5$. This is accompanied by a local electronic rearrangement within the FeAs layers. When
Ru doping approaches the critical threshold $x_c= 0.6$, corresponding to percolation of a magnetic square lattice
with nearest neighbor (n.n.) and next-nearest neighbor hopping, both magnetism and SC vanish.

\begin{figure}
\includegraphics[width=0.39\textwidth,angle=0]{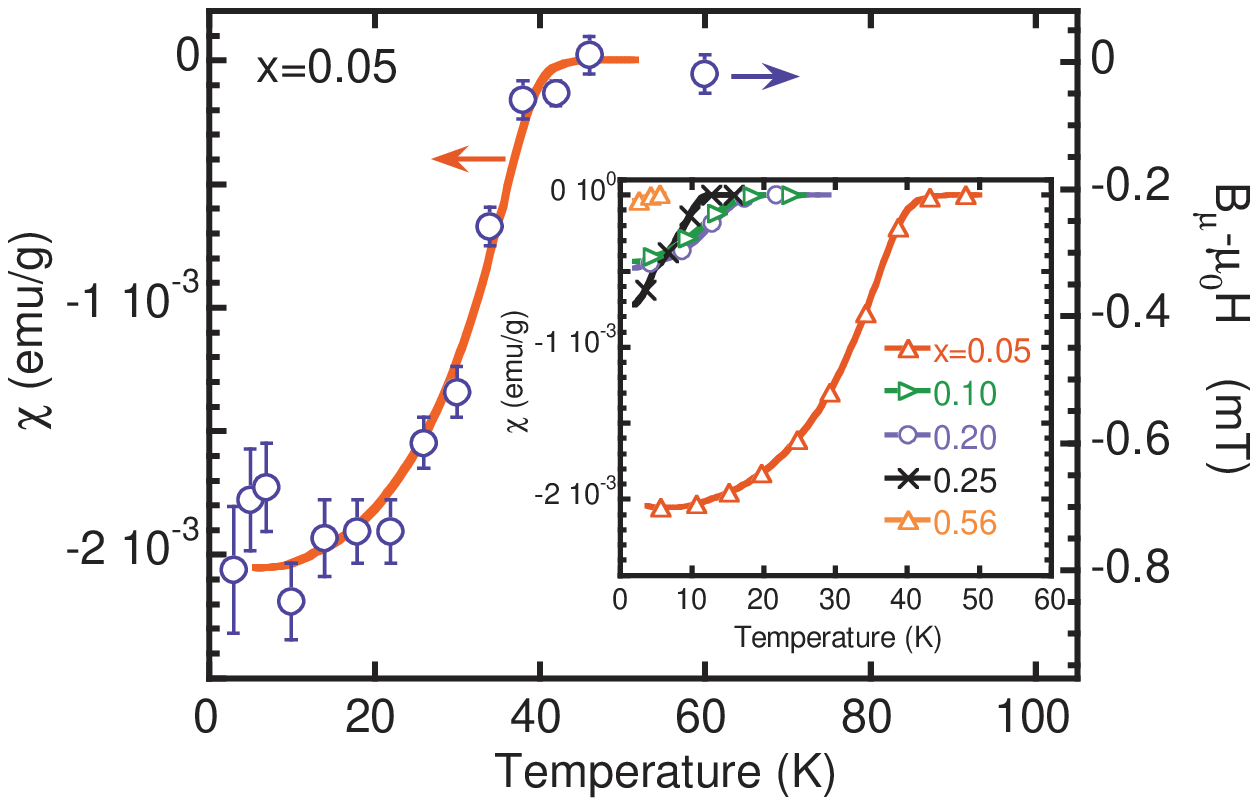}%
\caption{\label{fig:squid}(Color online) Muon diamagnetic shift (TF, $\mu_0H$=20 mT, {\color{Blue} $\circ$}) and ZFC SQUID susceptibility ($\mu_0H$=0.5 mT,
solid curve) for $x=0.05$. Inset: ZFC SQUID susceptibility $\chi$(T) of representative \smfraof\ samples ($\mu_0H$=0.5 mT). No relevant differences are found from FC measurements (not shown).}
\end{figure}

The investigated polycrystalline \smfraof\ samples are the same of Ref.~\onlinecite{Tropeano2010}. From $^{19}$F
nuclear magnetic resonance the relative fluorine content was found to be constant within $\Delta \lesssim0.01$ in the whole
set of samples investigated. To investigate the bulk character of the superconducting state we carried out
transverse field (TF)-$\mu$SR measurements, where a sample is field-cooled (FC) in a magnetic field larger than
the lower superconducting critical field $H_{c1}$, applied perpendicular to the initial muon-spin orientation ($\bm{H}\perp \bm{S}_\mu$).
A flux-line lattice (FLL) is thus generated below $T_c$ and the muon-spin precessions around the local field, $B_\mu$, display a diamagnetic
shift $B_\mu\!=\! \mu_0 H(1+N_e\chi)$, with $\chi<0$ and $N_e$ an effective demagnetization factor.
For samples with $x<0.1$ the time evolution of the full TF-$\mu$SR asymmetry is fitted to a single SC component as ${\cal A}_{TF}(t) \!\!=\!\! a_{TF} e^{-\sigma_{TF}^2 t^2/2}\cos(\gamma B_\mu t) $, where $a_{TF}$ is the amplitude, $\gamma\!=\!8514\,\mu$s$^{-1}$/T is the muon gyromagnetic ratio and $\sigma_{TF}$ the time decay of the precession. This fit holds over the entire $T$ range (not shown) as expected in a bulk superconductor. The shift of B$_\mu$ with respect to the applied field is displayed vs. temperature in Fig.~\ref{fig:squid} (symbols) for $x=0.05$, together with the rescaled susceptibility (solid curve).

By contrast, for $x\geq 0.1$ the muon relaxation (not shown) becomes extremely fast already above the superconducting $T_c$. In polycrystals this is the signature of magnetic order taking place at $T_M>T_c$ \cite{Allodi2006}. A coexisting FLL is hidden by the large magnetic damping, preventing its investigation by TF-$\mu$SR. We characterized the superconducting response by DC magnetic susceptibilities $\chi$(T) (inset of Fig.~\ref{fig:squid}), measured by a Superconducting Quantum Interference Device (SQUID). Comparison with the bulk case, $x=0.05$  allows us to roughly estimate the SC volume fraction of the other samples, which becomes absolutely marginal above $x=0.5$, confirming previous reports \cite{Tropeano2010}. The shielding response is instead observable, but already much weaker for all samples with a reduced $T_c$, at odds with the behavior of analogous Nd-1111 and La-1111 Ru-doped samples
\cite{Lee2010, Satomi2010}, suggesting that superconductivity is not a homogeneous bulk phenomenon in \smfraof\ for $x\geq0.1$.

\begin{figure}
\includegraphics[width=0.37\textwidth,angle=0]{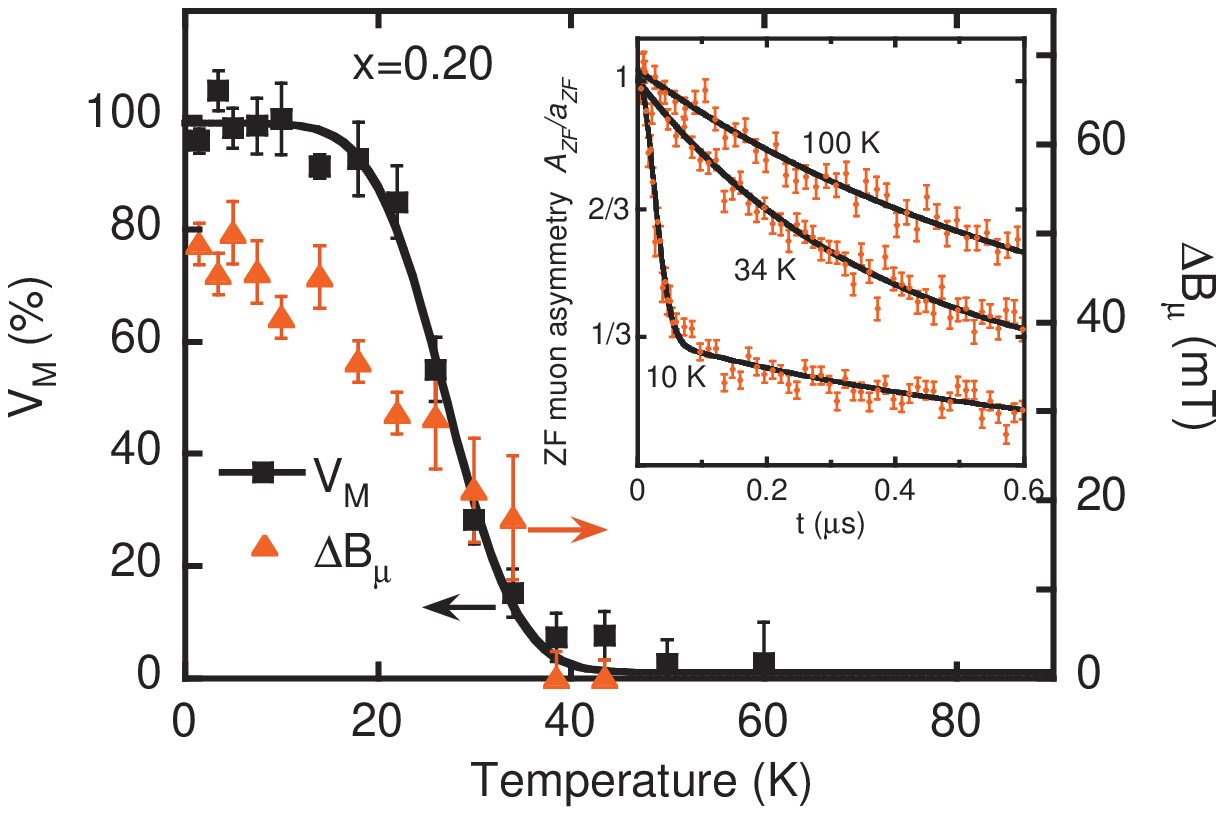}%
\caption{\label{fig:AF}(Color online)
Fraction of sample volume $V_{\mathrm{M}}$ ($\blacksquare$) where muons detect an internal magnetic field
and its width $\Delta
$B$_{\mu}$ ({\color{Red}$\blacktriangle$}) vs.~$T$, for of \smfraof\ with $x$=0.2. Inset: time dependence of the normalized ZF muon asymmetry, with best fits, at $T>T_M, T\lesssim T_M$ and $T\ll T_M$ }
\end{figure}

We now turn to zero-field (ZF)-$\mu$SR results, sensitive to short range magnetic order, owing to the short range muon-spin coupling with the electronic moments. The time evolution
of the ZF-$\mu$SR asymmetry is displayed in the inset of Fig.~\ref{fig:AF} for $x=0.20$ at three representative temperatures.
Below a mean transition temperature $T_M\approx30$ K the muon asymmetry is  ${\cal A}_{\mathrm{ZF}}(t) = a_{\mathrm{ZF}}(w_{\mathrm{T}} e^{-\sigma^2_{\mathrm{T}}
t^2/2} + w_{\mathrm{L}} \, e^{-\lambda_{\mathrm{L}} t})$  distinguishing a fast $\sigma_{\mathrm{T}}$ and a slow $\lambda_{\mathrm{L}}$ decay, where $a_{\mathrm{ZF}}$ is the high $T$ value and $w_{\mathrm{T}}+w_{\mathrm{L}}\!=\!1$. The two components reflect the orientation of the internal field with respect to the initial muon spin $\mathbf{S}_\mu$ (transverse, $\bm{B}_i\perp \bm{S}_\mu$, weight $w_{\mathrm{T}}\le 2/3$, and longitudinal, weight $w_{\mathrm{L}}$, $\bm{B}_i\!\parallel\! \bm{S}_\mu$ plus, possibly, a fraction with vanishing internal fields). The very fast transverse relaxation ($\sigma_{\mathrm{T}}\approx 30~\mu s^{-1}$) represents the signature of a sizeable distribution of internal fields $\bm{B_i}$ with standard deviation
$\Delta B_\mu\!=\!\sigma_{\mathrm{T}}/\gamma\!=\!(\overline{B_i}^2)^{1/2}$ reaching 45 mT at low $T$ (Fig.~\ref{fig:AF}). This is a typical Fe dipolar field value at the muon site in F-doped 1111 close to a M-SC crossover \cite{Sanna2009,Sanna2010}. The transverse component is overdamped down to 1.5 K (no asymmetry oscillations in Fig.~\ref{fig:AF}, inset) and its fast relaxation is partially quenched in fields of order $\overline{B_i}\approx 50$ mT applied along the initial muon-spin direction (not shown), as expected for a {\em static} $\bm{B}_i$. Therefore the overdamped muon-spin precessions are due to inhomogeneous short-range order \cite{Sanna2010}.

The fraction of the sample volume where muons experience a net field $\bm{B}_i$ is calculated as
$V_{\mathrm{M}}\!=\!3w_{\mathrm{T}}/2\!=\!3(1-w_{\mathrm{L}})/2$ \cite{Sanna2010}. Its dependence on temperature is shown
in Fig.~\ref{fig:AF} for $x=0.20$. It is noteworthy that $V_{\mathrm{M}}$ is 100\% already below
$20$ K, i.e. internal fields develop throughout the whole sample well above the ordering temperature of the
Sm sublattice ($T_N^{\mathrm{Sm}}\simeq5$ K) \cite{Maeter2009, Sanna2009} confirming  that the observed magnetic
order comes from the FeAs layers.
In summary, since in all samples with $x\geq0.10$ every muon detects a local magnetic field, every muon site must still be extremely close to some ordered magnetic moment. On the other hand in the same samples SQUID susceptibility detects a non negligible shielding fraction. This scenario is similarly found in cuprates and pnictides \cite{Keren2003,Sanna2009,Drew2009,Lang2010,Sanna2010} and it may originate from magnetic and superconducting interspersed regions of nanometric size, compatible with proximity and the very short coherence length $\xi$ of non-conventional superconductors \cite{Sanna2010}. Muons will measure a net field everywhere in this nanoscopic mixture of M and SC regions, if these three length-scales are comparable: the coherence length $\xi$, the mean SC domain size $d$ and the decay distance $r$ of the dipolar field $\bm{B}_i$ from the closest M cluster (typically a few nm).

\begin{figure}
\includegraphics[width=0.3\textwidth,angle=0]{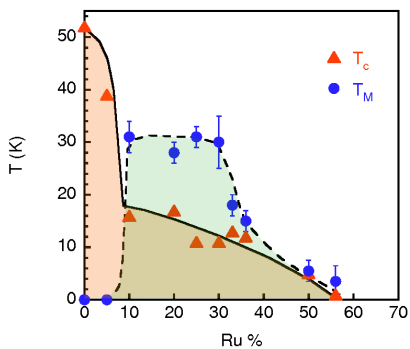}
\caption{\label{fig:phasediag}(Color online) a) SQUID superconducting T$_c$ ({\color{Red} $\blacktriangle$}) and ZF-\-$\mu$SR
magnetic T$_M$ ({\color{Blue} $\bullet$}) of \smfraof\ vs.~Ru content. The lines are guide for the eye.}
\end{figure}

The complete phase diagram for \smfraof, obtained by combining the SQUID and the ZF-$\mu$SR data, is shown in
Fig.~\ref{fig:phasediag}. At small Ru content a reentrant magnetic order within the FeAs layers is observed
at the same $x\sim 0.10$, where $T_c$ is dramatically reduced. Magnetic order in pnictides requires an orthorhombic structure, which has been recently demonstrated to occur \cite{Martinelli2011} below 150 K already in our starting material, SmFeAsO$_{0.85}$F$_{0.15}$. Figure \ref{fig:phasediag} shows also that, at high doping
levels, both M and SC states disappear around the same threshold
value $x_c\sim 0.6$, which is characteristic of {\em magnetic} interactions with nearest and next-nearest neighbors on a square lattice \cite{Papinuto2005}.
Notice that the same critical Ru content $x_c \approx 0.6$ is required to disrupt both superconductivity in \refraof\
with RE=La, Nd \cite{Lee2010, Satomi2010} and magnetic order in fluorine free \refrao\ with RE=La, Pr
\cite{McGuire2009, Bonf2011}. On one
hand this common critical Ru threshold emphasizes the intimate relation between magnetism and superconductivity. On the other
hand the degradation of SC when M is recovered indicates a strong competition between
the two respective order parameters. This seeming incongruity can be resolved by assuming that, in agreement with the most popular view, superconductivity is assisted by magnetic fluctuations. These fluctuations are suppressed by the full magnetic order, which competes with the SC state. Still short-range M order may allow fluctuations to survive alongside the M clusters, in nanoscopic form, and SC may survive as well. Upon approaching $x_c\approx0.6$ both the static magnetism and the relevant magnetic fluctuations vanish, together with the  superconductivity. Hence our results provide strong evidence for a superconductivity pairing of magnetic origin.

The presence of magnetic order in optimally F-doped RE1111 compounds was never observed before, and it is
very surprising that it is induced by Ru substitution, isoelectronic to Fe and diamagnetic
\cite{McGuire2009, Tropeano2010, Bonf2011}. This scenario bears some analogies with that observed in cuprates, where the isoelectronic and diamagnetic Cu:Zn substitution recovers a frozen spin configuration, producing a concomitant sharp suppression of superconductivity \cite{Alloul2009, Adachi2008}.

\begin{figure}
\includegraphics[width=0.45\textwidth,angle=0]{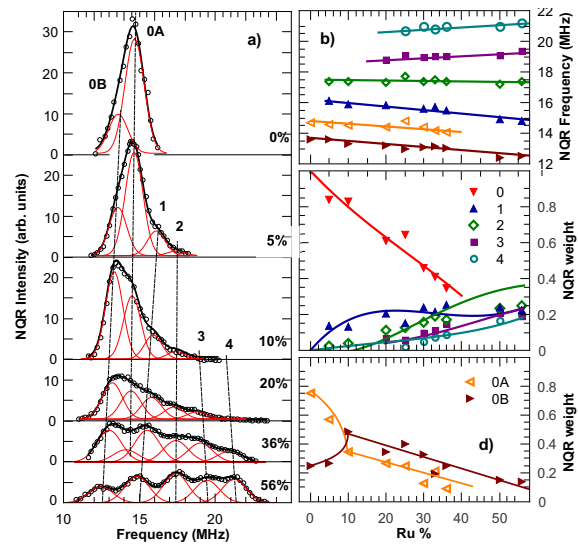}
\caption{\label{fig:NQRspe}(Color online) a) NQR spectra, for T=290 K, with best fits and their components for selected \smfraof\ samples; $x=0$, $T_c=52$ K, nominal F content 0.10, from Ref.\onlinecite{Lang2010}, NQR measured at $T=160$ K (the differences in nominal F and $T$ are irrelevant here). b) NQR peak frequencies vs.~Ru content. Lines guide the eye to identify n.n. Ru configurations. c) Normalized NQR weights, $a_j,~j=0\cdots4$ ($j=0$ is the sum of $0A$ and $0B$), vs. Ru content, with fits (solid, dash-dotted, see text). d) Weights of $0A, 0B$ ($j=0$ components).}
\end{figure}

Lang et al. \cite{Lang2010} have shown that in F-doped 1111 two As local environments appear, as witnessed by two NQR peaks.
They account for an intrinsic bimodal electronic inhomogeneity
on a nanoscopic scale within the FeAs layers, being associated \cite{Lang2010} to a charge-poor and a charge-rich local
environment, respectively. NQR maps the local electronic configuration via the electric-field gradient (EFG) at the $^{75}$As nucleus and we report here how its bimodal EFG is affected by Ru substitution.
In figure \ref{fig:NQRspe}a we show the NQR spectra for representative \smfraof\ samples, together with the $x=0$ data from Ref.~\onlinecite{Lang2010}.
In addition to the double peak structure (which is still visible, although much less pronounced than in F-underdoped samples \cite{Lang2010}), up to four new satellites appear, each characterized by the number of Ru n.n. ions in the local 4-fold coordination around As. Each spectrum has been fitted to a number of equal width Gaussian curves, varying from two to six. The five nearly equally-spaced peaks of the $x=0.56$ sample ($j=0 \div 4$) may be initially assigned to configurations with $j$ Ru n.n., respectively.

Figure \ref{fig:NQRspe}b displays the best-fit central
frequency of the different peaks as a function of Ru content. They can be easily grouped into six families indicated by the straight lines.
Figure \ref{fig:NQRspe}c displays the weight of these peaks as a function of Ru content, measured as the relative area $a_j=A_j/A_{\rm Tot}$ ($A_{\rm Tot}=\sum_j{A_j}$ is the total area), proportional to the number of $^{75}$As nuclei in that environment. The $j=0$ weight is taken as the sum of the two lower-frequency peaks, labelled $0A$ and $0B$. Indeed Fig.~\ref{fig:NQRspe}d shows their separate concentration dependence, with a remarkable initial correlation: one collapses while the other grows with increasing Ru content. This suggests them to be two components of the same $j=0$ Ru-free configuration. The assignment is supported by the good agreement of $a_j(x)$ with the fit to a binomial distribution, corrected with a weight redistribution that favors larger $j$, indicating a tendency towards Ru clustering (Fig.~\ref{fig:NQRspe}c solid lines).

\begin{figure}
\includegraphics[width=0.35\textwidth,angle=0]{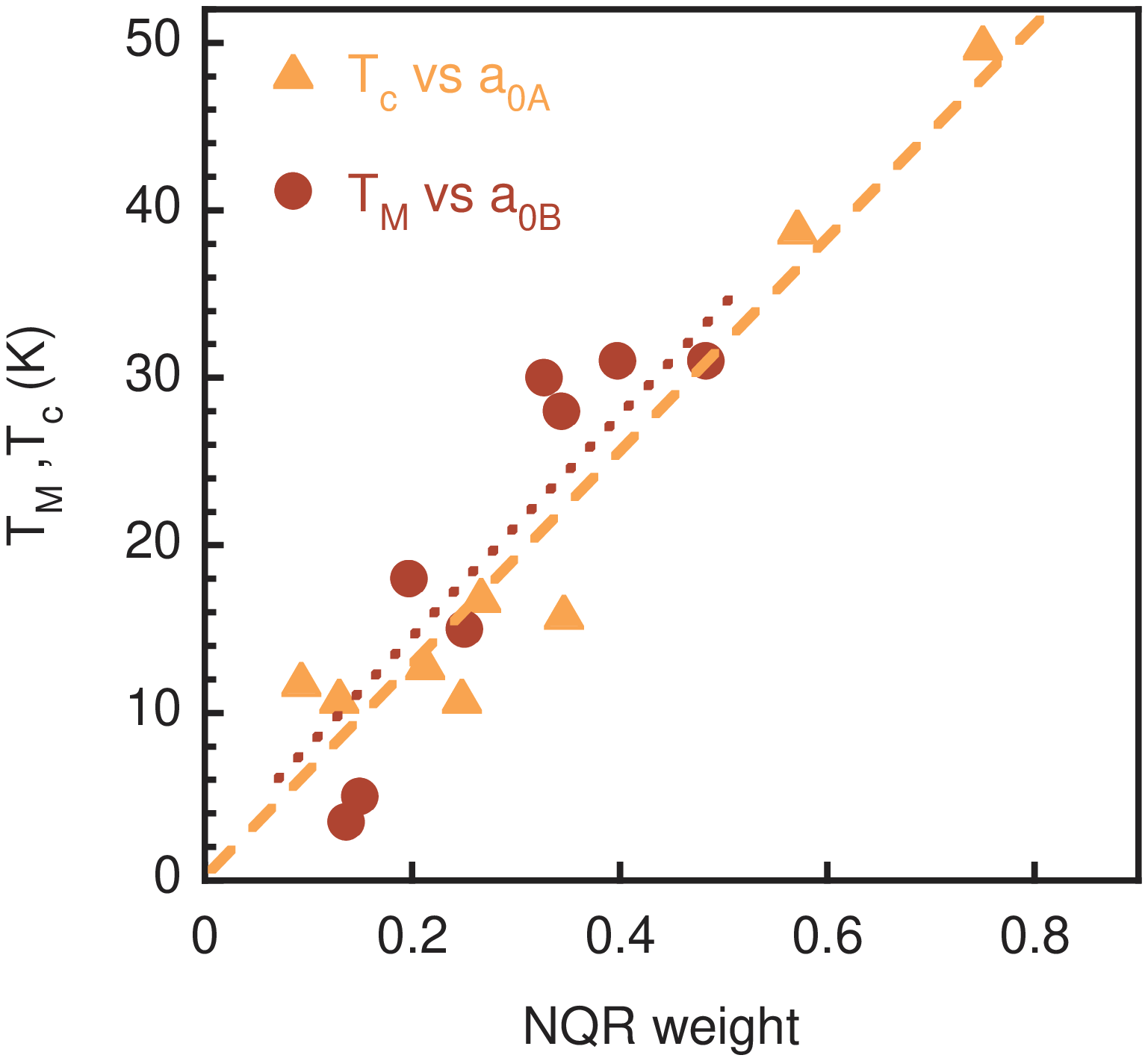}
\caption{\label{fig:TcNQR}(Color online)  T$_c$ and T$_M$ versus the NQR weights $a_{0A}$ and $a_{0B}$ respectively with their linear regressions.
}
\end{figure}

The average effect of Ru substitution on the electronic properties is directly reflected in the weight of the two low
frequency peaks, $a_{0A}, a_{0B}$, around 14.5 MHz and 13 MHz respectively, (Fig.~\ref{fig:NQRspe}b), representing nuclei without n.n. Ru ions. Comparison with the $x=0$ spectrum in Fig.~\ref{fig:NQRspe} clarifies that this is the same doublet already present at $x=0$ and discussed in Ref.~\cite{Lang2010} in terms of the electronic inhomogeneity of underdoped SC.
Their symmetric anti-correlated behavior for $x<0.1$, and the clear kinks for $x\approx 0.1$ are clearly related to the change in the magnetic and superconducting properties of the system.

This is confirmed in Fig.~\ref{fig:TcNQR},
showing a nearly linear dependence of both $T_c$ and $T_M$ when plotted vs.~the NQR weights $a_{0A}$ and $a_{0B}$, respectively.  This correlation provides compelling evidence that the volume fraction of the charge-rich (charge-poor) local environments in the FeAs layers increases together with the strength of the average SC coupling (M coupling), i.e the transition temperature $T_c$ ($T_M$). The relation between the average coupling and the spatial extension of the ordered phase suggests a percolation transition for both orders.
Given that the total charge doping is nearly constant as a function of isoelectronic Fe:Ru substitution in 1111 compounds \cite{McGuire2009,
Tropeano2010, Bonf2011}, the transition must be influenced directly by a  {\em local} charge redistribution due to Ru. The presence of Ru favors the charge-poor environment (weight $a_{0B}$) correlated with $T_M$, and
causes the decrease of the charge-rich volumes (weight $a_{0A}$) correlated with $T_c$. We notice incidentally that only the $j$=0 configuration shows the original doublet (i.e.~one n.n. Ru ion is enough to wipe out the intrinsic electronic inhomogeneity).

In conclusion, we have shown that the isoelectronic Fe:Ru substitution in optimally F-doped \smfraof\ leads to a re-entrant static magnetic order which degrades
the superconducting ground-state. The two order parameters compete, producing a nanoscopic phase separation. The two weights, tuned by the average Ru doping, scale with the corresponding transition temperatures. Both magnetism and superconductivity are suppressed at the percolation threshold characteristic of the magnetic system, suggesting that superconductivity cannot exist if magnetism is definitely suppressed by magnetic dilution. This picture strongly supports magnetic coupling models of superconductivity.

\textbf{Acknowledgments} $\mu$SR experiments were carried out at the PSI and ISIS muon facilities. We acknowledge partial financial support from the PRIN-08 project and from PSI EU funding. Thanks are due to M. Tropeano, A. Palenzona and C. Ferdeghini for the fruitful collaboration.


\end{document}